\documentstyle[prd,aps]{revtex}

\newcommand{\bea}{\begin{eqnarray}}
\newcommand{\eea}{\end{eqnarray}}

\begin{document}

\draft
\twocolumn[\hsize\textwidth\columnwidth\hsize\csname
@twocolumnfalse\endcsname

\title{$f(R)$ gravity theory and CMBR constraints} 
\author{Jai-chan Hwang${}^{(a)}$ and Hyerim Noh${}^{(b)}$}
\address{${}^{(a)}$ Department of Astronomy and Atmospheric Sciences,
                    Kyungpook National University, Taegu, Korea \\
         ${}^{(b)}$ Korea Astronomy Observatory,
                    San 36-1, Whaam-dong, Yusung-gu, Daejon, Korea \\
         }
\date{\today}
\maketitle

\begin{abstract}

We consider the large-scale cosmic structure generation by an
inflation based on a pure $f(R)$-type gravity theory.
Comparison with recent CMBR observations gives the following results:
(1) The near Zel'dovich spectral conditions uniquely choose $R^2$-type 
    gravity.
(2) The $R^2$ gravity predicts specific nearly scale-invariant 
    Zel'dovich spectra for both the scalar- and tensor-type perturbations.
Thus, 
(3) the considered model survives current observational data.
The {\it COBE}-DMR quadrupole data 
(4) give constraints on the coupling constant and the energy scale 
    during inflation, and
(5) require the gravitational wave contribution to be suppressed.
(6) Therefore, future observations of (2) and (5) can provide strong
    tests of the inflation scenario based on $R^2$ gravity.
Parallel analyses made in the conformally transformed Einstein frame give 
the observationally identical results.

\end{abstract}

\noindent

\vskip2pc]

\section{Introduction}
                                             \label{sec:Introduction}

The inflation scenario in the early universe provides a consistent and 
successful mechanism for generating the seed fluctuations of the observed 
large-scale cosmic structures as well as the gravitational wave background
\cite{inflation}.
As a natural consequence, the observed large-scale structures, in particular, 
the observed anisotropies of the cosmic microwave background radiation (CMBR) 
in the large-angular-scale in return, can provide constraints on the physics 
during the inflation era which is usually based on model scalar fields or 
generalized versions of gravity theories.
Recent spectacular developments in the CMBR anisotropy observations
\cite{Bennett-etal-1996,deBernardis-etal-2000,Hanany-etal-2000}
are transforming this branch of cosmology into a quantitative science.

In the theoretical side, the multipole moments of the CMBR anisotropies are 
basically governed by the seed fluctuations of the scalar- and tensor-type
structures which provide the initial spectra, and the transfer function
which incorporates the evolution of the initial spectra through 
recombination till the present epoch.
The transfer function is known to depend rather sensitively on many
cosmological parameters: present day Hubble constant, spatial curvature,
cosmological constant, density parameters of variety of constituents of 
the universe, possible reionization, etc. 
However, with the observational perspective at present epoch alone we
cannot distinguish the transfer function from the initial condition. 
In choosing the initial conditions one needs prior assumptions on the 
favored seed generating mechanism.
Although it is a common practice in the literature to assume a simple 
scale-invariant power-spectrum based on the single inflaton fluctuation
without tensor-type contribution, the initial conditions can 
be convoluted with the transfer functions in a rather intricated manner:
both encouraging \cite{Kanazawa-etal-2000} and discouraging
\cite{Bucher-etal-2000} conclusions can be drawn from reasonable variations
of the initial conditions.
Since different theoretical inflation models provide different seed 
generating mechanism it can affect only the initial conditions.

In the present work we investigate another theoretical aspect
of the study showing that the inflationary seed generating mechanism 
based on the completely different gravity theory can in fact give similar
viable initial conditions.
Here, we will consider a gravity theory which is a general function of the
scalar curvature.
Historically, the original proposal of a modern version of accelerated 
expansion in the early universe was, in fact, based on the gravity with 
quantum one-loop correction \cite{Starobinsky-1980}.
The curvature-square terms naturally appear in many theories: gravity 
with quantum corrections, Kaluza-Klein models, string/M-theory program, etc.
The $R^2$ gravity, in particular, has a natural reheating mechanism
\cite{Starobinsky-1980,Kofman-etal-1985,Starobinsky-1985,Vilenkin-1985,Kofman-Mukhanov-1986,Mijic-etal-1986,Kofman-etal-1987,Mukhanov-etal-1987}
thus providing a self-contained inflation model without introducing
a field or a phase transition.

In this paper, using the large-angular-scale CMBR data we will derive 
constraints on an inflation scenario based on $f(R)$ gravity theory.
We will take a phenomenological approach by trying to reconstruct
the successful inflation based on $f(R)$ gravity from observational results.
We consider general $f(R)$ term in the Lagrangian, and show that
the ordinary slow-roll assumption during inflation together with
Zel'dovich spectral conditions for the scalar- and tensor-type structures
chooses $R^2$ gravity as the unique candidate, eq. (\ref{pure-R^2}).
The considered models naturally lead to specific near scale-invariant
Zel'dovich spectra both for the scalar- and tensor-type perturbations,
eqs. (\ref{n-result-f},\ref{n-result}).
{}From the {\it COBE}-DMR data we derive a strong constraint on the
coupling constant of the $R^2$ term and the energy scale during
inflation epoch, eq. (\ref{result}).
It also predicts the the gravitional wave contribution should be
negligible, eq. (\ref{result}).
We show that parallel analyses made in the conformally transformed
Einstein frame give the observationally identical results.

Before presenting our main results in \S \ref{sec:SR}-\ref{sec:CT}, 
in \S \ref{sec:Classical}-\ref{sec:Observation} 
we summarize recent developments of the classical evolution
and the quantum generation processes for the scalar- and tensor-type
perturbations, and their observational consequences;
these are presented in a general context and in a unified manner.
We set $c \equiv 1$.

\section{Classical evolution}
                                             \label{sec:Classical}

We consider a gravity
\bea
   L = {1 \over 2} f (\phi, R) - {1 \over 2} \omega (\phi) \phi^{;c} \phi_{,c}
       - V (\phi),
   \label{Lagrangian}
\eea
where $f(\phi, R)$ is a general algebraic function of the scalar field $\phi$
and the scalar curvature $R$; $\omega (\phi)$ and $V (\phi)$ are
general algebraic functions of $\phi$.
We introduce $F \equiv \partial f / \partial R$.

We consider a spatially {\it flat} homogeneous and isotropic FLRW metric 
together with the most general scalar- and tensor-type spacetime 
dependent perturbations
\bea
   d s^2
   &=& - a^2 \left( 1 + 2 \alpha \right) d \eta^2
       - a^2 \beta_{,\alpha} d \eta d x^\alpha
   \nonumber \\
   & & + a^2 \left( \delta_{\alpha\beta}
       + 2 \varphi \delta_{\alpha\beta} + 2 \gamma_{,\alpha\beta}
       + 2 C_{\alpha\beta} \right) d x^\alpha d x^\beta,
   \label{metric-general}
\eea
where $C_{\alpha\beta}$ is a transverse-tracefree
tensor-type perturbation corresponding to the gravitational wave.
We also decompose $\phi = \bar \phi + \delta \phi$,
$F = \bar F + \delta F$, etc.

In \cite{GGT-H,GGT-HN,GGT-Q,GGT-scalar,GGT-GW} we analysed the classical 
evolution and the quantum generation processes of the scalar- and 
tensor-type structures.
Perturbed action can be written in a unified form as 
\cite{GGT-scalar,GGT-GW}
\bea
   \delta^2 S = {1 \over 2} \int a^3 Q \left( \dot \Phi^2
       - {1 \over a^2} \Phi^{|\gamma} \Phi_{,\gamma} \right) dt d^3 x.
   \label{action-unified}
\eea
This is valid for the second-order gravity system such as
either $f = F(\phi) R$ in the presence of a field $\phi$ or
$f = f(R)$ without the field; an overdot indicates the time derivative 
based on $t$ with $dt \equiv a d \eta$.
{}For the scalar- and tensor-type perturbations we have \cite{GGT-HN}:
\bea
   & & \Phi = \varphi_{\delta \phi}, \quad
       Q = { \omega \dot \phi^2 + 3 \dot F^2 / 2 F
       \over ( H + \dot F / 2 F )^2 },
   \nonumber \\
   & & \Phi = C^{\alpha}_{\beta}, \quad \; Q = F,
   \label{Phi-def}
\eea
where $H \equiv \dot a/a$.
$\varphi_{\delta \phi} \equiv \varphi - H \delta \phi/\dot \phi$ is
a gauge-invariant combination introduced in \cite{Lukash-1980,Mukhanov-1988}.
The equation of motion and the solution in the large-scale limit become
\bea
   & & {1 \over a^3 Q} (a^3 Q \dot \Phi)^\cdot - {\Delta \over a^2} \Phi = 0,
   \label{Phi-eq} \\
   & & \Phi = C ({\bf x}) - D ({\bf x}) \int_0^t {dt \over a^3 Q}.
   \label{Phi-sol}
\eea
Thus, ignoring the transient solution we have a temporally conserved behavior
\bea
   \Phi ({\bf x}, t) = C ({\bf x}).
\eea
Using $z \equiv a \sqrt{Q}$ and $v \equiv z \Phi$ eq. (\ref{Phi-eq}) becomes 
$v^{\prime\prime} + \left( k^2 - {z^{\prime\prime} / z} \right) v = 0$,
where a prime denotes the time derivative based on $\eta$.
Thus, the large-scale assumption used in eq. (\ref{Phi-sol})
is $z^{\prime\prime}/z \gg k^2$.

\section{Quantum generation}
                                             \label{sec:Quantum}

We introduce parameters \cite{GGT-HN}
\bea
   & & \epsilon_1 \equiv {\dot H \over H^2}, \quad
       \epsilon_2 \equiv {\ddot \phi \over H \dot \phi}, \quad
       \epsilon_3 \equiv {1 \over 2} {\dot F \over H F}, \quad
       \epsilon_4 \equiv {1 \over 2} {\dot E \over H E},
   \label{epsilons-def}
\eea
where $E \equiv F [ \omega + 3 \dot F^2 / (2 \dot \phi^2 F) ]$.
{}For $\dot \epsilon_1 = 0$ we have $\eta = - 1 / [(1 + \epsilon_1) aH]$,
thus, $a \propto |\eta|^{-1/(1 + \epsilon_1)} \propto t^{-1/\epsilon_1}$.
{}For $\dot \epsilon_i = 0$ we have $z^{\prime\prime} / z = n / \eta^2$
where for the scalar- ($n = n_s$) and tensor-type ($n = n_t$) perturbations
\cite{GGT-HN,GGT-GW}
\bea
   & & n_s = { ( 1 - \epsilon_1 + \epsilon_2 - \epsilon_3 + \epsilon_4 )
       ( 2 + \epsilon_2 - \epsilon_3 + \epsilon_4 )
       \over ( 1 + \epsilon_1 )^2 },
   \nonumber \\
   & & n_t = { ( 1 + \epsilon_3 ) ( 2 + \epsilon_1 + \epsilon_3 )
       \over (1 + \epsilon_1)^2 }.
   \label{n}
\eea

With constant $n$ eq. (\ref{Phi-eq}) becomes a Bessel's equation.
After a canonical quantization we can derive the
vacuum expectation value of the quantum field $\hat \Phi$.
In the quantization process of the gravitational wave we need to take
into account of the two polarization states properly
\cite{Ford-Parker-1977,GGT-GW}.
In the large scale limit, $k|\eta| \ll 1$, the power-spectra based on 
vacuum expectation value become \cite{GGT-scalar,GGT-GW}
\bea
   & & {\cal P}^{1/2}_{\hat \varphi_{\delta \phi}} (k, \eta)
       = {1 \over \sqrt{Q}} {H \over 2 \pi} {1 \over a H |\eta|}
       {\Gamma (\nu_s) \over \Gamma (3/2)}
       \left( {k |\eta|\over 2} \right)^{3/2 -\nu_s},
   \label{P-LS-scalar} \\
   & & {\cal P}^{1/2}_{\hat C_{\alpha\beta}} (k, \eta)
       = \sqrt{2 \over Q} {H \over 2 \pi} {1 \over a H |\eta|}
       {\Gamma (\nu_t) \over \Gamma (3/2)}
       \left( {k |\eta|\over 2} \right)^{3/2 - \nu_t},
   \label{P-LS-GW}
\eea
where $\nu_s \equiv \sqrt{n_s + 1/4}$ and $\nu_t \equiv \sqrt{n_t + 1/4}$,
and we have {\it ignored} dependences on the vacuum choices;
for $\nu_s = 0 = \nu_t$ we have additional $\ln{(k|\eta|)}$ factors.

\section{Observational consequences}
                                             \label{sec:Observation}

The observationally relevant scales exit Hubble horizon within about
$60$ $e$-folds before the end of the latest inflation.
{}Far outside the horizon the quantum fluctuations classicalize
and we can identify ${\cal P}_{\Phi} = {\cal P}_{\hat \Phi}$ where
the l.h.s. is the power-spectrum based on spatial averaging.
Since $\Phi$ is conserved while in the large-scale limit, 
the power-spectra in eqs. (\ref{P-LS-scalar},\ref{P-LS-GW})
can be identified as the classical power-spectra at later epoch.
Later, we will consider a scenario with the inflation based on $f(R)$
gravity followed by ordinary radiation and matter dominated eras
based on Einstein gravity.
We have shown in \cite{GGT-H} that as long as the scale remains in the 
super-horizon scale $\Phi$ is conserved independently of the changing gravity 
theory from one type to the other.
Thus, since eqs. (\ref{P-LS-scalar},\ref{P-LS-GW}) are now valid for 
the classical power-spectra
the spectral indices of the scalar and tensor-type perturbations are
\bea
   n_S - 1 = 3 - \sqrt{ 4 n_s + 1 }, \quad n_T = 3 - \sqrt{ 4 n_t + 1 }.
   \label{n-spectra}
\eea

{}For the scale independent Zel'dovich ($n_S -1 = 0 = n_T$) spectra
the quadrupole anisotropy becomes \cite{HN-PRL}
\bea
   \langle a_2^2 \rangle
   = \langle a_2^2 \rangle_S + \langle a_2^2 \rangle_T
   = {\pi \over 75} {\cal P}_{\varphi_{\delta \phi}}
       + 7.74 {1 \over 5} {3 \over 32} {\cal P}_{C_{\alpha\beta}}.
   \label{a_2}
\eea
The ratio between two types of perturbations is \cite{Knox} 
\bea
   r_2 \equiv {\langle a_2^2 \rangle_T \over \langle a_2^2 \rangle_S}
       = 3.46 { {\cal P}_{C_{\alpha\beta}} \over 
       {\cal P}_{\varphi_{\delta F}} }.
   \label{r-def}
\eea
The four-year {\it COBE}-DMR data give \cite{COBE}:
\bea
   \langle a_2^2 \rangle \simeq 1.1 \times 10^{-10}.
   \label{a_2-value}
\eea

\section{Spectral constraints}
                                            \label{sec:SR}

We consider a gravity 
\bea
   L = {1 \over 2} f (R),
   \label{Lagrangian-f}
\eea
which is a special case of eq. (\ref{Lagrangian}).
Results in \S \ref{sec:Classical}-\ref{sec:Observation}
remain valid by setting $\omega \dot \phi^2$ term 
in eq. (\ref{Phi-def}) equal to zero, and replacing $\varphi_{\delta \phi}$ 
to $\varphi_{\delta F} \equiv \varphi - H \delta F/\dot F$.
Consequently, the slow-roll parameter $\epsilon_4$ in eq. (\ref{epsilons-def})
changes to $\epsilon_4 \equiv \ddot F / (H \dot F)$ and we have 
$\epsilon_2 = 0$ \cite{GGT-HN}.

Equations for the background are \cite{GGT-H,GGT-HN}
\bea
   & & H^2 = {1 \over 3F} \left( {RF - f \over 2} - 3 H \dot F \right),
   \label{BG1} \\
   & & \dot H = - {1\over 2 F} \left( \ddot F - H \dot F \right),
   \label{BG2}
\eea
with $R = 6 ( 2 H^2 + \dot H )$.
Equation (\ref{BG2}) gives 
\bea
   \epsilon_1 = \epsilon_3 ( 1 - \epsilon_4).
   \label{epsilon_34}
\eea
Since we have $R = 6 ( 2 + \epsilon_1 ) H^2$, eq. (\ref{BG1}) gives
$f = R F (1 + \epsilon_1 - 2 \epsilon_3) / (2 + \epsilon_1)$.

As the slow-roll conditions, as in the MSF (minimally coupled scalar field) 
\cite{Stewart-Lyth-1993}, we consider
\bea
   |\epsilon_1| \ll 1.
   \label{slow-roll}
\eea
We have $\epsilon_2 = 0$.
Equations (\ref{P-LS-scalar},\ref{P-LS-GW}) are available for 
$\dot \epsilon_i = 0$; since the large-scale structures are generated from 
a short duration (about 60 $e$-folds) of the inflation we expect time 
variation of $\epsilon_i$ during that period is negligible.
The spectral indices in eqs. (\ref{n},\ref{n-spectra}) are valid for 
arbitrary but constant $\epsilon_i$s.
{}For constant $\epsilon_i$ we have
$f \propto R^{(2 + \epsilon_1)/(1 + \epsilon_1 - 2 \epsilon_3)}$.
{}From eqs. (\ref{epsilon_34},\ref{slow-roll}) we have 
$\epsilon_3 \simeq 0$ or $\epsilon_4 \simeq 1$.
If $\epsilon_4 \simeq 1$, the spectral constraint $n_S - 1 \simeq 0$ 
thus $n_s \simeq 2$ in 
eq. (\ref{n}) implies $\epsilon_3 \simeq 1$ or $4$ both of which are excluded
if we impose $n_T \simeq 0$ thus $n_t \simeq 2$ in eq. (\ref{n}).
Thus we have $\epsilon_3 \ll 1$ and
\bea
   f \propto R^{2 - \epsilon_1 + 4 \epsilon_3}.
   \label{fR-slow-roll}
\eea
Imposing $\epsilon_3 \simeq 0$ in eq. (\ref{n}) with $n_s \simeq 2$
leads to $\epsilon_4 \simeq 0$ or $-3$; correspondingly we have 
$\epsilon_1 \simeq \epsilon_3$ or $4 \epsilon_3$.
In the case with $\epsilon_4 \simeq -3$ we have exactly
$f = R^2$ to the linear order in $\epsilon_i$!
In the case with $\epsilon_4 \simeq 0$ we have
$\epsilon_3 = \epsilon_1$, thus
\bea
   f \propto R^{2 + 3 \epsilon_1}.
   \label{pure-R^2}
\eea
In this case we have $|\epsilon_i| \ll 1$.

Therefore, the requirement of near scale-invariant Zel'dovich spectra for 
{\it both} structures generated from inflation based on $f(R)$ gravity
chooses $f \propto R^2$ gravity.
The current observations of CMBR, however, provide a constraint on only the 
sum of the two types of structures. 
It is often believed that the observations provide constraint on the
scalar-type perturbation, thus its amplitude and the spectral index.
In such a case, our use of the Zel'dovich type spectral index for the
tensor-type structure as well to sort out the $R^2$ can be regarded as
having a loop-hole in the argument.
However, notice that we have used the $n_T$ constraint to exclude 
the cases of $\epsilon_4 \simeq 1$, thus $\epsilon_3 \simeq 1$ or $4$
just above eq. (\ref{fR-slow-roll}).
If we accept these cases, we have $f \propto R^{-2}$ or $R^{-2/7}$,
respectively, and eqs. (\ref{n},\ref{n-spectra}) give $n_T \simeq -2$ or $-8$, 
respectively, which look like not very promising cases.

In the case of $|\epsilon_i| \ll 1$ the spectral indices in 
eq. (\ref{n-spectra}) become
\bea
   n_S - 1 \simeq 2 ( 3 \epsilon_1 - \epsilon_4 ), \quad
       n_T \simeq 0,
   \label{n-result-f}
\eea
to $\epsilon_i$-order.
We can estimate $\epsilon_4 = 3 \epsilon_1 + \dot \epsilon_1/(H \epsilon_1)$,
and because we have $\dot \epsilon_1 \sim {\cal O} (\epsilon_i^2)$
we cannot ignore the second term.
In the MSF case eqs. (\ref{n},\ref{n-spectra}) remain valid with
$\epsilon_3 = 0 = \epsilon_4$, thus
we have $n_S - 1 \simeq 4 \epsilon_1 - 2 \epsilon_2$ and
$n_T \simeq 2 \epsilon_1$.

\section{COBE-DMR constraints}
                                             \label{sec:R^2}

Now, we consider the following gravity during inflation 
\bea
   f = {m_{pl}^2 \over 8 \pi} \left( R + {R^2 \over 6 M^2} \right).
   \label{Lagrangian-R^2}
\eea
We can show that the trace of gravitational field equation
	 gives $\Box R - M^2 R = 0$ \cite{GGT-H}. 
	 Thus, $R$ behaves like a MSF with mass $M$.
There have been many studies of the $R^2$ inflation model
\cite{Starobinsky-1980,Starobinsky-1981,Starobinsky-1983,Kofman-Mukhanov-1986,Mijic-etal-1986,Kofman-etal-1987,Mukhanov-etal-1987,Morris-1989,Mukhanov-1989,Salopek-etal-1989,Baibosunov-etal-1990,Gottlober-etal-1992}.
During inflation we {\it assume} the second term dominates over the
Einstein action which {\it requires} $H^2_{\rm infl} \gg M^2$;
this is a slow-roll regime and we have $|\epsilon_i| \ll 1$.
{}For the model in eq. (\ref{pure-R^2}), the pure $R^2$ gravity implies 
$\epsilon_1 = 0$ and exponential inflation.
However, in our case of eq. (\ref{Lagrangian-R^2}) 
we do not have vanishing $\epsilon_1$.
In the slow-roll regime we have
$\epsilon_3 = \epsilon_1$, thus eq. (\ref{n-result-f}) remains valid.

{}From eqs. (\ref{P-LS-scalar},\ref{P-LS-GW}) we have
\bea
   {\cal P}^{1/2}_{\varphi_{\delta F}} = {H \over 2 \pi} {1 \over \sqrt{Q}},
       \quad
       {\cal P}^{1/2}_{C_{\alpha\beta}} = {H \over 2 \pi} \sqrt{2 \over Q}.
\eea
Assuming $H^2 \gg M^2$ we have
\bea
   {\cal P}^{1/2}_{\varphi_{\delta F}}
       = {1 \over 2 \sqrt{3 \pi}} {1 \over |\epsilon_1|} {M \over m_{pl}}, 
       \quad
       {\cal P}^{1/2}_{C_{\alpha\beta}}
       = {1 \over \sqrt{\pi}} {M \over m_{pl}},
   \label{P-spectra}
\eea
where we ignored $[ 1 + {\cal O} (\epsilon_i)]$ factors.
Thus, ${\cal P}^{1/2}_{C_{\alpha\beta}}/{\cal P}^{1/2}_{\varphi_{\delta F}}
= 2 \sqrt{3} |\epsilon_1|$, and from eq. (\ref{r-def}) we have
\bea
   r_2 = 41.6 \epsilon_1^2.
   \label{r}
\eea
These results can be compared with the MSF case where the power-spectra 
ratio becomes $2 \sqrt{-\epsilon_1}$ and the quadrupole ratio 
$r_2 = - 6.93 n_T$ which is the well known consistency relation;
for a consistency relation with general $\Lambda$, see \cite{Knox-1995}.
Since $r_2$ in eq. (\ref{r}) is quadratic order in $\epsilon_1$, and 
$n_T$ in eq. (\ref{n-result-f}) vanishes to the linear order in $\epsilon_i$,
in order to check whether the consistency relation holds in the $R^2$ gravity
we need to investigate the second-order effects of the slow-roll parameters.

{}From eqs. (\ref{a_2},\ref{P-spectra}) using eq. (\ref{a_2-value}) we have
\bea
   {M \over m_{pl}} = 3.1 \times 10^{-4} |\epsilon_1|.
   \label{M-constraint}
\eea
Similar constraint has been known in the literature
\cite{Kofman-Mukhanov-1986,Mijic-etal-1986,Kofman-etal-1987,Mukhanov-etal-1987,Salopek-etal-1989,Gottlober-etal-1992}.
Compared with previous works which are based on an asymptotic solution
of the background $R^2$-gravity inflation model (see below), 
our approach is based on exact solutions available when $\dot \epsilon_i = 0$, 
where instead we can handle the perturbations
analytically; also, our results are derived without using the
conformal transformation.

In fact, based on the known inflation solution we can make the constraints
more concrete.
{}From eqs. (\ref{BG1},\ref{BG2}) we can derive
$ \dot H^2 - 6 H^2 \dot H - H^2 M^2 - 2 H \ddot H = 0$.
In the limit $H^2 \gg M^2$, and in the regime of vanishing $\ddot H$ term, 
we have a solution during the inflation as $H = H_i - M^2 (t - t_i)/6$ 
\cite{Kofman-etal-1987}.
Assuming that the inflation ends near $(t_e - t_i) \sim 6 H_i/M^2$, 
we can estimate the number of $e$-folds from $t_k$ (when the relevant scale 
$k$ exits the Hubble horizon and reaches the large-scale limit)
till the end of inflation as 
\bea
   N_k \equiv \int_{t_k}^{t_e} H dt = - {H^2 \over 2 \dot H} (t_k) 
       = - {1 \over 2 \epsilon_1 (t_k)}.
   \label{N_k}
\eea
[In this case we are considering piecewise constancy of $\epsilon_1$.
         Since $\dot \epsilon_1 \sim {\cal O} (\epsilon_i^2)$,
         eqs. (\ref{P-LS-scalar},\ref{P-LS-GW}) are not affected
         to the linear order in slow-roll parameters.]
Thus $H(t_k)/M = \sqrt{N_k / 3}$.
Since the large-angular CMBR scales exit horizon about
$50 \sim 60$ $e$-folds before the end of the latest inflation,
$N_k \sim 60$ requires $|\epsilon_1| = {1 \over 2 N_k} \sim 0.0083$.
Consequently from eqs. (\ref{r},\ref{M-constraint}) we have
\bea
   r_2 \sim 0.0029, \quad
      {M \over m_{pl}} \sim 2.6 \times 10^{-6}, \quad
      {H(t_k) \over M} \sim 4.5.
   \label{result}
\eea
Thus we have negligible gravitational wave contribution to CMBR anisotropies.
In addition, we can show that $\dot \epsilon_1 = - 2 \epsilon_1^2 $
and $\ddot H \sim {\cal O} (\epsilon_1^3)$, thus
$\epsilon_4 = \epsilon_1$ and eq. (\ref{n-result-f}) becomes
$n_S - 1 \simeq 4 \epsilon_1 \sim - {2 / N_k}$, thus
\bea
   n_S - 1 \simeq - 0.033, \quad n_T \simeq 0.00.
   \label{n-result}
\eea

\section{Conformal transformation method}
                                             \label{sec:CT}

It is well known that by a conformal transformation the $f(R)$ gravity 
in eq. (\ref{Lagrangian-f}) can be transformed to Einstein gravity plus 
a minimally coupled scalar field with a special potential 
\cite{CT,Hwang-CT-1997}.
{}For $R^2$ gravity in eq. (\ref{Lagrangian-R^2}) 
under the conformal transformation $\hat g_{ab} \equiv \Omega^2 g_{ab}$
with $\Omega \equiv \sqrt{8 \pi F}/m_{pl}
\equiv e^{{1 \over 2} \sqrt{16 \pi \over 3} \phi/m_{pl}}$, thus
$\phi \equiv \sqrt{3 \over 16 \pi} m_{pl}\ln{ (1 + {1 \over 3 M^2} R) }$,
the potential becomes
\bea
   V(\phi) = {3M^2 m_{pl}^2 \over 32 \pi} \left( 1 
       - e^{- \sqrt{16 \pi / 3} \phi/m_{pl}} \right)^2.
\eea
In the limit $\phi/m_{pl} \gg 1$ (corresponding to $R \gg M^2$) we have 
a soll-roll inflation, whereas
in the other limit $\phi \ll m_{pl}$ we have $V = {1 \over 2} M^2 \phi^2$
which provides a reheating; $R^2$ gravity has both features simultaneously
as well \cite{Kofman-etal-1987}.
The two gravity theories conformally related are not necessarily the same 
physically.
The equations for background are:
\bea
   H^2 = {8 \pi \over 3 m_{pl}^2} \left( {1 \over 2} \dot \phi^2 + V \right), 
       \quad
       \ddot \phi + 3 H \dot \phi + V_{,\phi} = 0.
\eea

Now, we derive the spectra generated during the slow-roll inflation regime.
The number of $e$-folds in eq. (\ref{N_k}) gives
\bea
   N_k = {8 \pi \over m_{pl}^2} \int_{\phi_e}^{\phi_k}
       {V d \phi \over V_{,\phi} }
       = {3 \over 4} \sqrt{3 \over \pi} { H^2 m_{pl} \over V_{,\phi} } (t_k).
\eea
We can estimate $\epsilon_i$'s in eq. (\ref{epsilons-def}) as
\bea
   \epsilon_1 = - {3 \over 4 N_k^2}, \quad
       \epsilon_2 = {1 \over N_k} + {3 \over 4 N_k^2}.
   \label{epsilons-CT}
\eea
The spectral indices are given by eqs. (\ref{n-spectra}) as
\bea
   n_S - 1 = - {2 \over N_k} - {9 \over 2 N_k^2}, \quad
       n_T = - {3 \over 2 N_k^2}.
   \label{n-CT}
\eea
The power-spectra follow from eqs.  (\ref{P-LS-scalar},\ref{P-LS-GW}) as
\bea
   {\cal P}^{1/2}_{\varphi_{\delta \phi}}
       = {1 \over \sqrt{3 \pi}} {M \over m_{pl}} N_k,
       \quad
       {\cal P}^{1/2}_{C_{\alpha\beta}}
       = {1 \over \sqrt{\pi}} {M \over m_{pl}},
    \label{P-spectra-CT}
\eea
where we ignored $[ 1 + {\cal O} (\epsilon_i)]$ factors.
These results {\it coincide exactly} with the expressions in 
eq. (\ref{P-spectra}):
this is understandable because both $\varphi_{\delta \phi}$
and $C^{(t)}_{\alpha\beta}$ are invariant under the conformal
transformation \cite{Hwang-CT-1997}.
We can show 
\bea
   r_2 = 3.46 {3 \over N_k^2} = - 6.92 n_T,
   \label{r_2-CT}
\eea
where the first step coincides with eqs. (\ref{r},\ref{N_k});
the second step is the well known consistency relation.
{}From eqs. (\ref{a_2},\ref{a_2-value},\ref{P-spectra-CT}) we have
\bea
   {M \over m_{pl}} = {1.6 \times 10^{-4} \over N_k}
       \sim 2.6 \times 10^{-6},
\eea
which coincides exactly with eq. (\ref{M-constraint}).
Therefore, using $N_k \sim 60$ the {\it observationally verifiable results}
of $n_S$, $n_T$ and $r_2$ in eqs. (\ref{n-CT},\ref{r_2-CT}) 
{\it coincide precisely} with the ones based on the original $R^2$ gravity: 
these are predictions on the gravitational wave contribution and the 
spectral index in eqs. (\ref{result},\ref{n-result}).

We can show that using the conformal transformation in 
\cite{Hwang-CT-1997} the results in the original frame
can be recovered from the ones derived in Einstein frame;
this is understandable because the two gravity theories are related to 
each other through conformal transformation in the action level.
In comparison, what we have shown above is that the observationally verifiable 
results in the two frames coincide exactly which, we believe, 
are not necessarily obvious.

\section{Discussions}
                                             \label{sec:Discussions}

The main results of this paper are 
eqs. (\ref{pure-R^2},\ref{result},\ref{n-result}).
We have two testable predictions of the inflation model based on successful
$f(R)$ gravity theory, thus $R^2$ gravity.
Firstly, we should have the specific nearly scale-invariant Zel'dovich spectra
in eq. (\ref{n-result}) which are valid considering the linear order 
in slow-roll parameters, i.e., ${\cal O} (\epsilon_i)$.
The recent high precision measurements of the CMBR by Boomerang and 
Maxima-1 in small angular scale \cite{deBernardis-etal-2000,Hanany-etal-2000}
together with {\it COBE}-DMR data \cite{Bennett-etal-1996} suggest the 
scalar spectral index $n_S -1 \simeq 0.01^{+0.17}_{-0.14}$ at the $95\%$ 
confidence level \cite{Jaffe-etal-2000}.
Thus, the inflation model considered in this paper can provide
viable seeds for the large-scale cosmic structures which pass
the current observational limits.
Secondly, in order to have a successful inflation we should have negligible
contribution of the gravitational wave to the CMBR anisotropies
as in eq. (\ref{result}).
The results in the conformally transformed Einstein frame provide
observationally identical results with precise
predictions on the gravitational wave contribution and the spectral index:
compare eqs. (\ref{n-CT},\ref{r_2-CT}) with
eqs. (\ref{result},\ref{n-result}).
Therefore, future observations of the spectral indices and/or the contribution 
of the gravitational wave to CMBR temperature/polarization anisotropies
will provide strong tests for the considered inflation model.
Other inflation models based on different gravity theories with similar 
scale-invariant Zel'dovich spectra and suppressed gravitational wave 
contribition are known in the literature \cite{HN-PRL}.

\section*{Acknowledgments}

We thank Dominik Schwarz for useful discussions during CAPP2000 conference 
in Verbier which initiated this work.
We also wish to thank Ewan Stewart for useful comments.
This work was supported by Korea Research Foundation Grant (KRF-99-015-DP0443).


\end{document}